\documentclass[twocolumn,prb,amsmath,amssymb,floatfix,superscriptaddress,showpacs]{revtex4-1}
\usepackage{graphicx}
\usepackage{dcolumn}
\usepackage{bm}
\usepackage{xspace}
\usepackage{color}
\def\cto{Cu$_{3}$TeO$_{6}$\xspace}

\begin{document}

\title{Discovery of coexisting Dirac and triply degenerate magnons in a three-dimensional antiferromagnet}
\author{Song~Bao}
\altaffiliation{These authors contributed equally to this work.}
\author{Jinghui~Wang}
\altaffiliation{These authors contributed equally to this work.}
\author{Wei~Wang}
\altaffiliation{These authors contributed equally to this work.}
\author{Zhengwei~Cai}
\author{Shichao~Li}
\author{Zhen~Ma}
\author{Di~Wang}
\author{Kejing~Ran}
\author{Zhao-Yang~Dong}
\affiliation{National Laboratory of Solid State Microstructures and Department of Physics, Nanjing University, Nanjing 210093, China}
\author{D.~L.~Abernathy}
\affiliation{Neutron Scattering Division, Oak Ridge National Laboratory, Oak Ridge, Tennessee 37831, USA}
\author{Shun-Li~Yu}
\email{slyu@nju.edu.cn}
\author{Xiangang~Wan}
\email{ccmp@nju.edu.cn}
\author{Jian-Xin Li}
\email{jxli@nju.edu.cn}
\author{Jinsheng Wen}
\email{jwen@nju.edu.cn}
\affiliation{National Laboratory of Solid State Microstructures and Department of Physics, Nanjing University, Nanjing 210093, China}
\affiliation{Collaborative Innovation Center of Advanced Microstructures, Nanjing University, Nanjing 210093, China}

\begin{abstract}
\noindent
Topological magnons are emergent quantum spin excitations featured by magnon bands crossing linearly at the points dubbed nodes, analogous to fermions in topological electronic systems. Experimental realization of topological magnons in three dimensions has not been reported so far. Here, by measuring spin excitations (magnons) of a three-dimensional antiferromagnet \cto with inelastic neutron scattering, we provide direct spectroscopic evidence for the coexistence of symmetry-protected Dirac and triply degenerate nodes, the latter involving three-component magnons beyond the Dirac-Weyl framework. Our theoretical calculations show that the observed topological magnon band structure can be well described by the linear-spin-wave theory based on a Hamiltonian dominated by the nearest-neighbour exchange interaction $J_1$. As such, we showcase \cto as an example system where Dirac and triply degenerate magnonic nodal excitations coexist, demonstrate an exotic topological state of matter, and provide a fresh ground to explore the topological properties in quantum materials.
\end{abstract}

\maketitle
\noindent {\bf Introduction}\\
By introducing the concept of topology into electronic bands, plenty of novel quantum materials, such as topological insulators\cite{RevModPhys.82.3045,RevModPhys.83.1057} with the edge state existing in the bulk gap, and Dirac\cite{PhysRevLett.108.140405,PhysRevB.85.195320,Liu21022014,liu2014cdas} and Weyl semimetals\cite{PhysRevB.83.205101,xu2015discovery,lv2015observation} featured by linear-band crossings at the Dirac and Weyl nodes, respectively, have been discovered. Excitations associated with these topological states are fermions described by the Dirac-Weyl equations\cite{wehling2014dirac,RevModPhys.88.021004}. Recently, exotic new fermions\cite{PhysRevX.6.031003,Bradlynaaf5037,nature547_298,PhysRevLett.119.206402}, such as the triply degenerate ones beyond such a classification, have emerged\cite{np14_349,nature546_627}, enriching the family of topological materials and advancing the understanding on band topology. Since topological band structure is independent of the statistics of the constituent quasiparticles, many efforts have been devoted to seeking for nontrivial topological analogues of fermions in bosonic systems, {\it e.g.}, phononic\cite{PhysRevLett.95.155901,PhysRevLett.117.068001,nature555_342,nature555_346,PhysRevLett.120.016401} and photonic crystals\cite{npho8_821,Lu622,nature553_59,Zhou1009,Yang1013,Hararieaar4003,Bandreseaar4005}. In two dimensions, various topological states for magnons (also bosons), which are spin excitations in magnetically ordered systems, have also been proposed. These include topological magnon insulators\cite{PhysRevB.87.144101,PhysRevLett.115.147201,PhysRevB.90.024412,nc6_6805,np13_736}, and magnonic Dirac\cite{PhysRevB.94.075401,PhysRevLett.119.107205,2399-6528-1-2-025007,PhysRevX.8.011010} and Weyl semimetals\cite{PhysRevLett.113.047202,nc7_12691,PhysRevLett.117.157204,PhysRevB.95.224403,PhysRevX.1.021002}. 
Following the successful examples in fermionic systems\cite{Bradlynaaf5037,nature547_298,np14_349}, triply degenerate nodal excitations have been predicted for the magnonic case\cite{PhysRevB.97.094412,0295-5075-120-5-57002}, extending the topological classification in bosonic systems. Topological magnonic systems exhibit: i) non-zero Berry curvature which gives rise to the anomalous Hall effect of the heat current carrying by the charge-neutral spin excitations\cite{PhysRevLett.104.066403,Onose297,PhysRevLett.106.197202,PhysRevB.85.134411,1367-2630-18-10-103039,nc6_6805,PhysRevLett.115.106603}; ii) edge or surface state that is topologically protected\cite{PhysRevB.87.144101,PhysRevB.87.174427,PhysRevB.90.024412,PhysRevB.97.081106}.
These exotic properties make the materials appealing in developing high-efficiency and low-cost spintronic devices\cite{RevModPhys.90.015005,Onose297,PhysRevB.97.081106,chumak2015magnon}. However, candidate materials to realize topological magnons are scarce. Especially for topological magnons in three dimensions, there has been no experimental report so far. In this regard, Li {\it et al.}\cite{PhysRevLett.119.247202} have predicted \cto to host Dirac magnons, offering an excellent opportunity for experimental investigations into the topological properties of magnons. 

In this work, we measure the spin excitations in \cto with inelastic neutron scattering (INS), and compare the INS data with the linear-spin-wave calculations performed based on a Hamiltonian dominated by the nearest-neighbour exchange interaction $J_1$. From the results, we discover symmetry-protected three-dimensional Dirac and triply degenerate magnons in \cto. 

\begin{figure}[htb]
\centering
\includegraphics[width=0.98\linewidth]{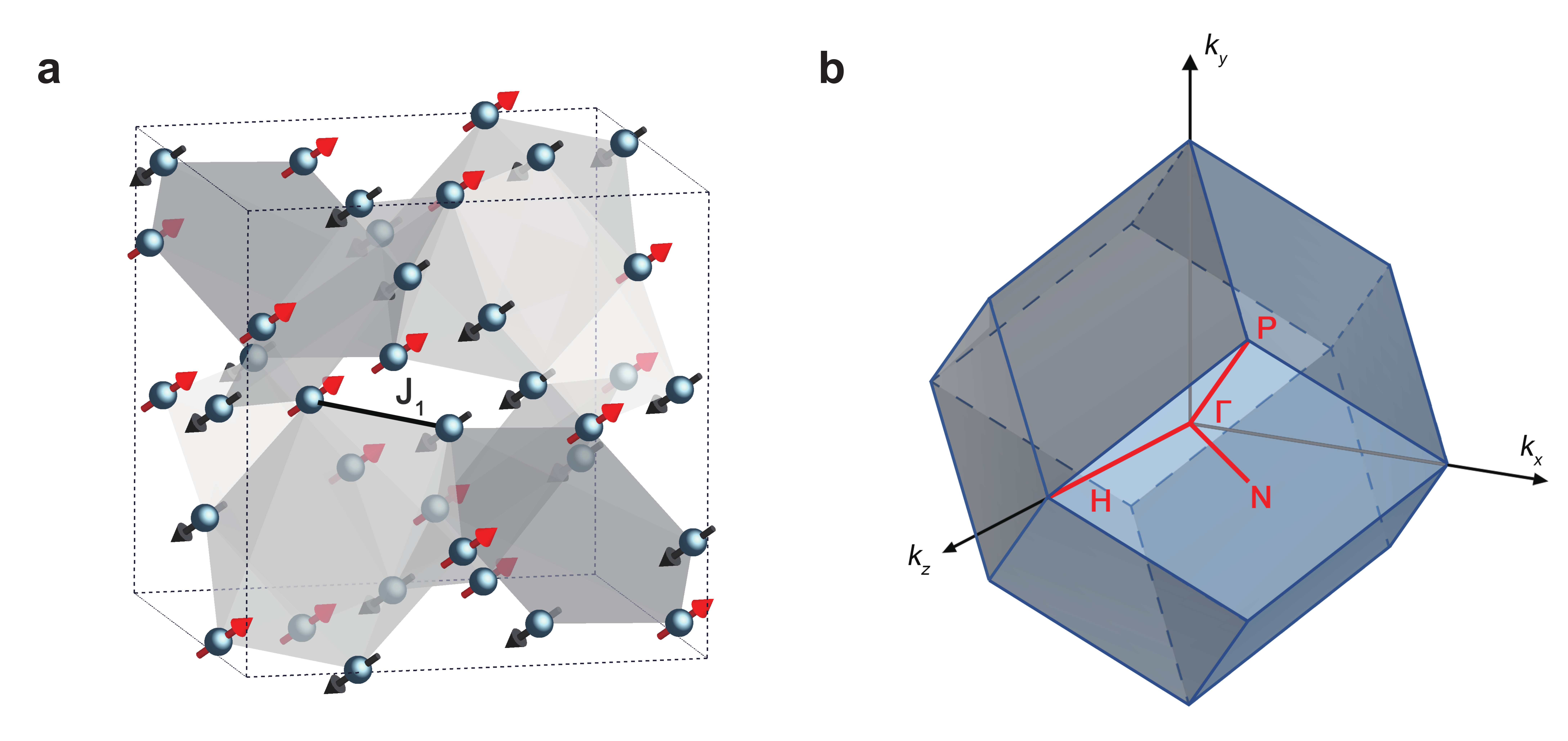}
\caption{\label{fig:structure}{Crystal structure and the first Brillouin zone for Cu$_{3}$TeO$_{6}$. {\bf a}, Schematic for the centro-symmetric cubic crystal structure with the $Ia$-3 space group (No.~206). For simplicity, only Cu$^{2+}$ ions with spins indicated by arrows are shown. Shades indicate hexagons formed by Cu$^{2+}$ ions. The nearest-neighbour exchange interaction $J_1$ is indicated by a solid line. {\bf b}, The first Brillouin zone of the primitive unit cell with high-symmetry paths and points.}}
\end{figure}

\begin{figure*}[htb]
\centering
\includegraphics[width=\linewidth]{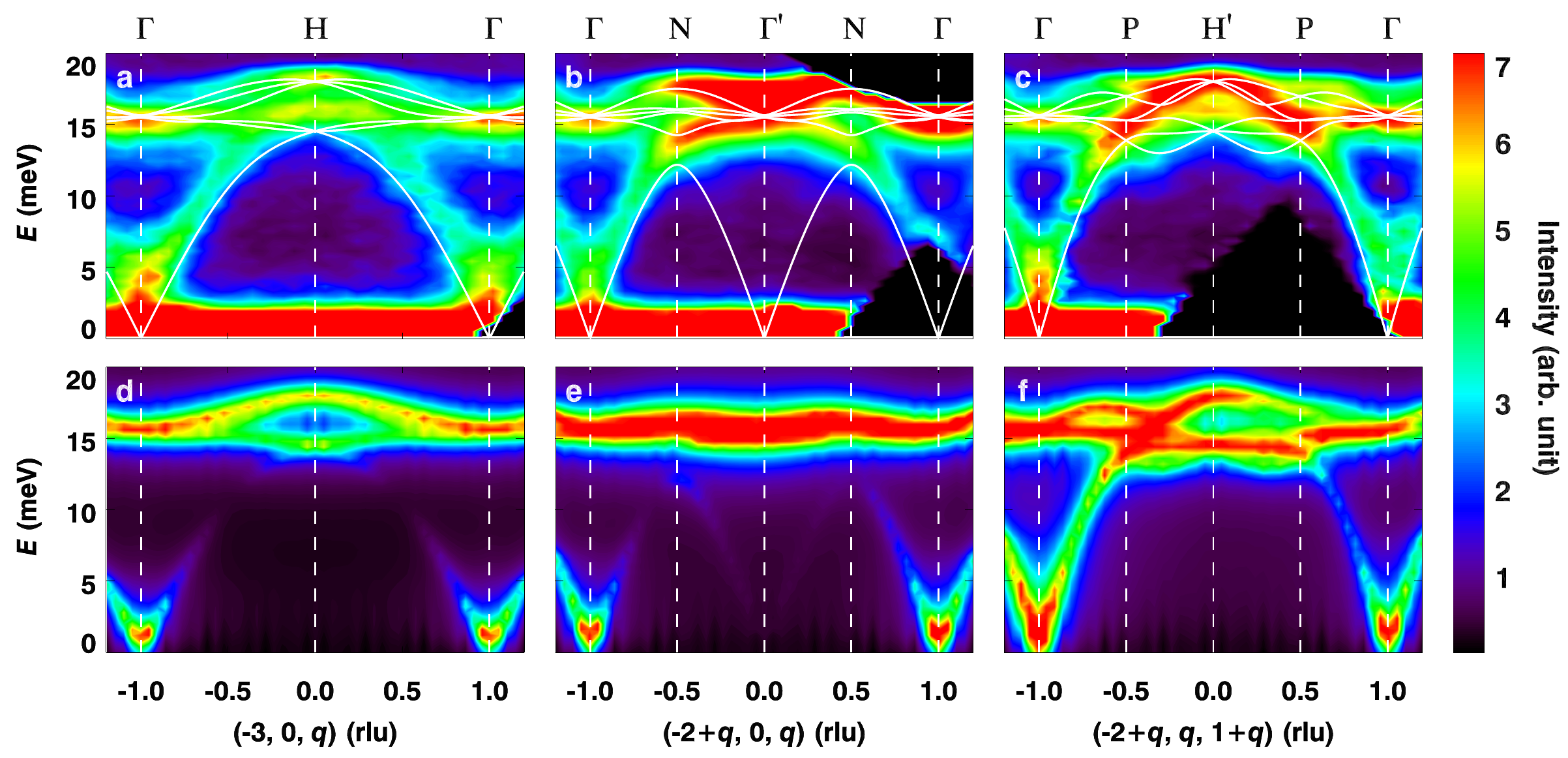}
\caption{\label{fig:dispersion}{Experimental and calculated spin-wave excitations. {\bf a}, {\bf b}, and {\bf c}, Inelastic neutron scattering results of the spin excitation spectra measured at $T=5$~K along [001], [101], and [111] directions, respectively. {\bf d-f}, Calculated magnetic spectra using the linear-spin-wave theory based on a set of parameters with $J_1=9.07$, $J_2=0.89$, $J_3=-1.81$, $J_4=1.91$, $J_5=0.09$, and $J_6$=1.83~meV. The uncertainties of the parameters are about 6\%. The calculated dispersions are plotted as solid lines in {\bf a-c}. The wave vector {\bf Q} is expressed as {\bf Q}=$(2\pi/a,\,2\pi/b,\,2\pi/c)$ reciprocal lattice unit~(rlu) with $a=b=c=9.537$(3)~\AA. Vertical dashed lines indicate the {\bf Q} positions illustrated in Fig.~\ref{fig:structure}{\bf b}.}}
\end{figure*}

\begin{figure}[htb]
\centering
\includegraphics[width=0.98\linewidth]{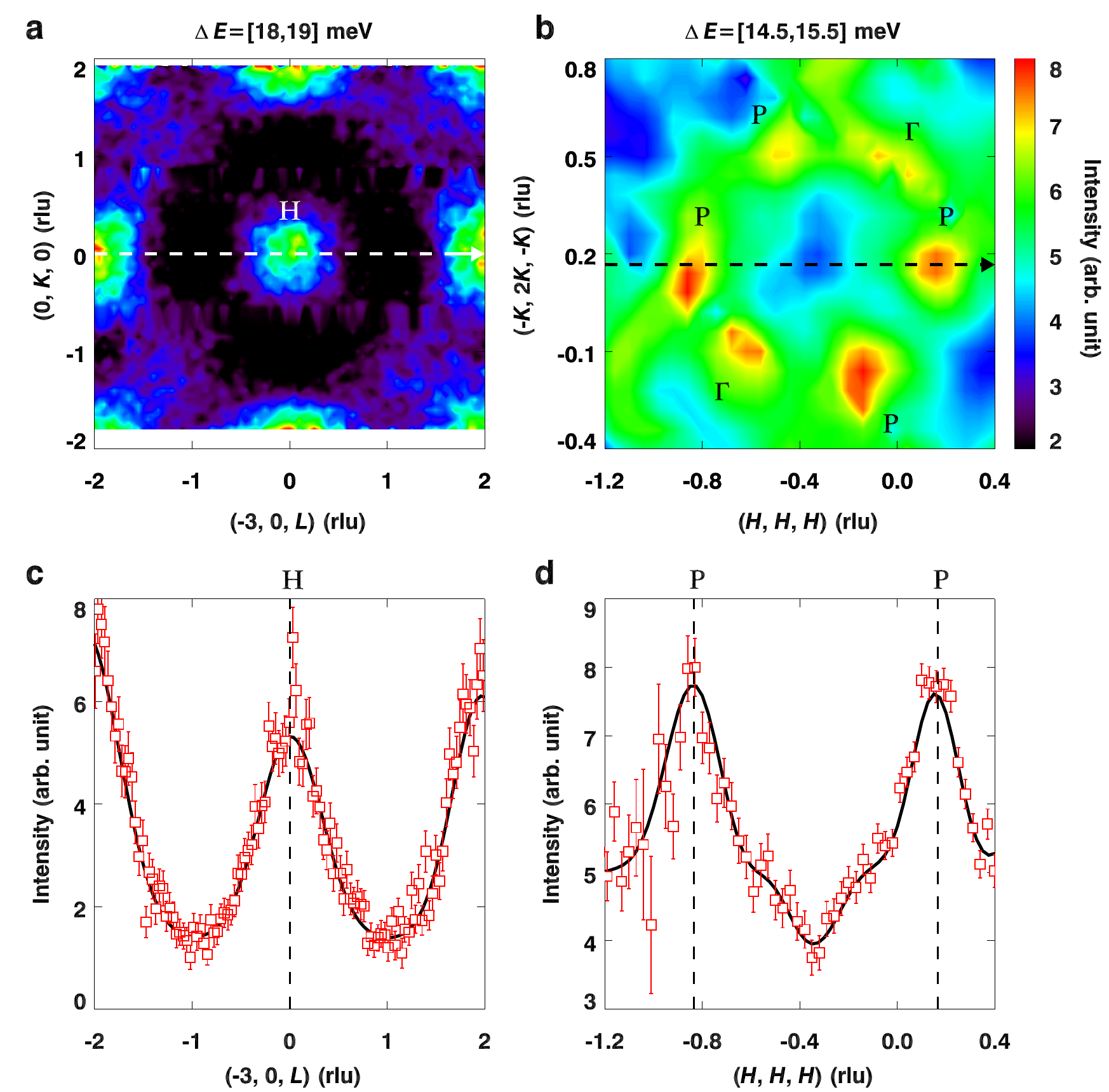}
\caption{\label{fig:ecuts}{Triply degenerate and Dirac nodes. {\bf a}, Contour plotted against two orthogonal axes [010] and [001] with an energy interval of $18.5\pm0.5$~meV, and {\bf b}, against [111] and $[\bar{1}2\bar{1}]$ with an energy interval of $15\pm0.5$~meV. Dashed arrows in {\bf a} and {\bf b} indicate the trajectories of the cuts plotted in {\bf c} and {\bf d}. Vertical dashed lines in {\bf c} and {\bf d} denote the triply degenerate and Dirac nodes, respectively. Lines through data are fits with Gaussian functions. Errors represent one standard deviation throughout the paper.}}
\end{figure}

\begin{figure*}[htb]
\centering
\includegraphics[width=0.88\linewidth]{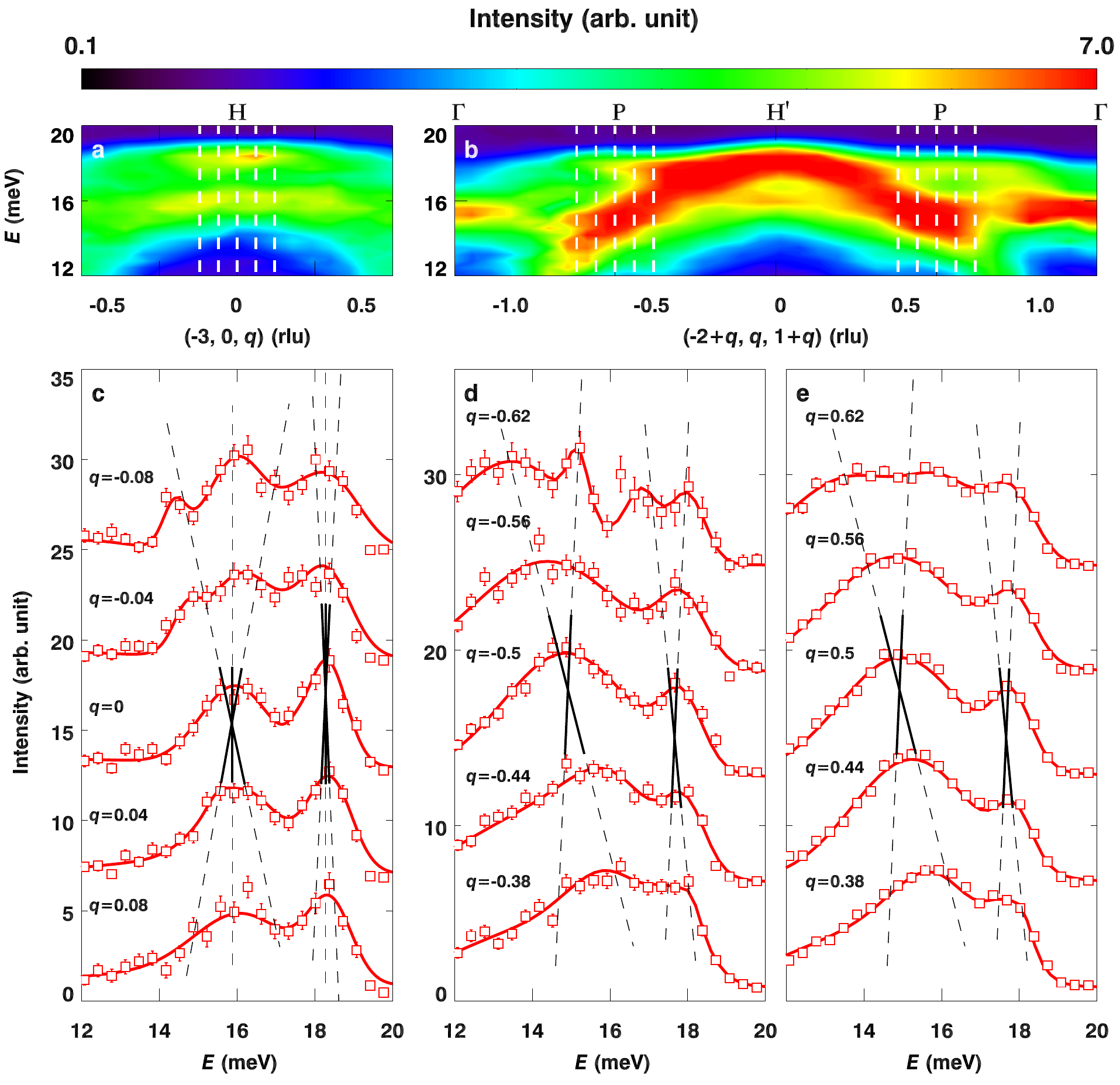}
\caption{\label{fig:qcuts}{Linear dispersions near triply degenerate and Dirac nodes. {\bf a}, Dispersions along the [001], and {\bf b}, [111] directions. {\bf c}, Constant-{\bf Q} (wave vector) cuts indicated by the vertical dashed lines in {\bf a}. Constant-{\bf Q} cuts along negative and positive $q$s in {\bf b} are plotted in {\bf d} and {\bf e}, respectively. The intensities of the cuts in {\bf c-e} are offset to elaborate the dispersion. Red solid lines through data are fits with Gaussian functions. Black solid lines in {\bf c-e} are guides to the eye illustrating the linearity of the dispersions, while dashed lines are their extensions.}}
\end{figure*}

\bigskip
\noindent {\bf Results}\\
\noindent{\bf Sample characterisations.} The crystal structure of \cto with the $Ia$-3 space group (No.~206) is illustrated in  Fig.~\ref{fig:structure}a~(ref.~\onlinecite{0953-8984-17-48-017}). Six Cu$^{2+}$ ions form an almost coplanar hexagon, and each ion is vertex-shared by two hexagons, constituting a three-dimensional spin-web structure\cite{0953-8984-17-48-017,MANSSON2012142,HE2014146,NORMAN2018507}. Neutron powder diffraction has shown that \cto develops a long-range collinear antiferromagnetic order below the transition temperature $T_{\rm N}$ of 61~K, with spins aligned along the [111] direction\cite{0953-8984-17-48-017,MANSSON2012142}. Such a magnetic state as illustrated in Fig.~\ref{fig:structure}a, with each up spin related to a down spin by centro-inversion, belongs to a magnetic group with the PT symmetry\cite{PhysRevLett.119.247202}, where P and T are space-inversion and time-reversal operations, respectively. Under the protection of this symmetry, magnons are expected to exhibit nontrivial topological properties\cite{PhysRevLett.119.247202}. Inelastic neutron scattering is a direct approach to visualize magnon bands in the momentum and energy space, which acts as angle-resolved-photoemission spectroscopy in characterising electronic band structures\cite{Liu21022014,xu2015discovery,nature546_627}. Below, we present results from INS measurements on well-characterised high-quality single crystals of \cto~(See Supplementary Figures~1 and~2 for details).

\noindent{\bf Magnetic excitation spectra.} We have obtained a rich INS dataset which covers up to 8 Brillouin zones in the whole energy range of interest at various temperatures. In Fig.~\ref{fig:dispersion}a, b, and c, we present the excitation spectra obtained at $T=5$~K along three high-symmetry directions [001], [101], and [111], respectively. These directions are illustrated in Fig.~\ref{fig:structure}b. The INS spectra show clear excitations dispersing up from the magnetic Bragg peaks\cite{0953-8984-17-48-017}. More magnetic peaks are shown in Supplementary Figure~4. These peaks can be almost perfectly indexed with the collinear magnetic structure which respects the PT symmetry\cite{0953-8984-17-48-017,MANSSON2012142}. We have performed measurements at higher temperatures up to $T=70$~K, above the $T_{\rm N}$ of 61~K, and the results are shown in Supplementary Figures~3 and~4. We find that the well-defined dispersions at $T=5$~K become almost featureless at 70~K~(Supplementary Figure~3g-i), along with the disappearance of the magnetic Bragg peaks (Supplementary Figure~4e-f). The wave-vector and temperature dependences of the excitations are clearly evidencing that they are spin-wave excitations. 

Turning back to the data at 5~K, we can see that the acoustic bands extend up to about 15~meV, and the optical bands are present roughly between 15 and 20~meV. As there are twelve Cu$^{2+}$ atoms in a primitive unit cell\cite{0953-8984-17-48-017}, there are six doubly degenerate magnon bands due to the PT symmetry\cite{PhysRevLett.119.247202}. Since these six bands coexist in such a narrow energy window, band crossings are expected. In fact, by bare visual inspection of Fig.~\ref{fig:dispersion}a-c, we can already identify various high-symmetry points at which the bands cross each other. Taking Fig.~\ref{fig:dispersion}a as an example, $\Gamma$ points at $E\approx15$~meV, and H point at $E\approx16$ and 18.5~meV exhibit as hot spots in the dispersion. At the H point, the interval between the two spots is clearly visible. To better characterise these points, we perform theoretical calculations as described below.

\medskip
\noindent{\bf Comparison with linear-spin-wave calculations.} The well-defined acoustic modes and quick disappearance of the magnetic order and excitations when approaching $T_{\rm N}$~(Supplementary Figure~4) indicate that \cto is a three-dimensional antiferromagnet without much frustration, consistent with a small frustration index of $f=|{\it\Theta}_{\rm CW}|/T_{\rm N}\approx2.9$ in our sample, where ${\it\Theta}_{\rm CW}=-175$~K is the Curie-Weiss temperature~(Supplementary Figure~2a). Therefore, we carry out the linear-spin-wave calculation to fit the experimental data. We find that a $J_1$-$J_2$ model with only nearest-neighbour (NN) and next-nearest-neighbour (NNN) exchange interactions cannot fit the data, given the apparent discrepancies especially on the optical branches~(Supplementary Figure~3a-f). We have added longer-range exchange interactions and found that at least up to sixth-NN~($J_6$) can we fit the data satisfactorily. The necessity for using terms up to $J_6$ may lie in the highly-interconnected three-dimensional spin network and the large number of Cu$^{2+}$ ions in the unit cell, such that differences between different exchange paths can be small. However, $J_1$ is the only dominant term, which is compatible with the modest frustration of the system as there is no comparable interaction to compete with $J_1$. The calculated spin-wave spectra using these parameters along [001], [101], and [111] directions are presented in Fig.~\ref{fig:dispersion}d, e and f, respectively, and the corresponding dispersions are plotted on top of the experimental data in Fig.~\ref{fig:dispersion}a-c. 

The calculated magnetic excitation spectra capture most of the features in the experimental results, as shown in Fig.~\ref{fig:dispersion}. We note that we can include more longer-range interactions to improve the fittings, mostly for the acoustic branches in the low-energy range where no band crossings occur. But given the present agreement between the theoretical and experimental results, we believe that our Hamiltonian up to $J_6$ is appropriate, since the main purpose for the calculations is to guide our characterisations on the crossing points, which are in the high-energy range. The comparison between the calculated dispersions and experimental data assures that we have observed multiple nodal points along different trajectories in the Brillouin zone in the energy range of 15 to 18.5~meV. Again, we remind that these nodes are symmetry protected\cite{PhysRevLett.119.247202}. The presence of nodes along all these directions indicates that the associated nodal excitations are of three-dimensional nature, similar to those in the fermionic systems\cite{Liu21022014,xu2015discovery}. We first identify the four-fold degenerate Dirac nodes, at which two doubly degenerate magnon bands cross each other, for example, the $\Gamma$($\Gamma^\prime$) and P points. 

In addition to the Dirac nodes predicted in ref.~\onlinecite{PhysRevLett.119.247202}, we also observe some double-triply (triply degenerate, hereafter) nodal points at some high-symmetry positions, {\it e.g.}, the H(H$^\prime$) points in Fig.~\ref{fig:dispersion}a(c). At the $\Gamma$ point, there exist both Dirac and triply degenerate nodes very close in energy. In fact, the triply degenerate nodes are generically expected for a system with a PT plus some point-group symmetry such as $C_3$~(refs~\onlinecite{Bradlynaaf5037,nature546_627,PhysRevB.97.094412,0295-5075-120-5-57002}), which is the case in \cto~(refs~\onlinecite{PhysRevLett.119.247202,0953-8984-17-48-017}). 

The existence of above mentioned nodal points is guaranteed by the symmetry, independent of the Hamiltonian\cite{PhysRevLett.119.247202,PhysRevB.97.094412,0295-5075-120-5-57002}. As a demonstration, we show the calculated results using a $J_1$-$J_2$ model in Supplementary Figure~3a-c---given the apparent failure of this model in describing the experimental data, this model still gives Dirac nodes at P and triply degenerate nodes at H. This further strengthens our conclusion that the unprecedented magnon band structure with coexisting Dirac and triply-degenerate nodes has been discovered in \cto. Since the $\Gamma$ point hosts two types of nodes too close in energy to be resolved experimentally, we pick the P and H points for further elaborations.

\bigskip
\noindent{\bf Triply degenerate and Dirac nodes.} In the dispersions shown in Fig.~\ref{fig:dispersion}a, we observe two triply degenerate points at H at two energies of $E\approx16$ and 18.5~meV. We first perform a constant-energy ($E$) cut in the dispersions at an energy interval of $\Delta E=18.5\pm0.5$~meV through the H point. The results are plotted in Fig.~\ref{fig:ecuts}a, where it clearly shows a circle centring at the H point. A cut along the [001] direction through this point yields a peak exactly centring at the H point, as shown in Fig.~\ref{fig:ecuts}c. Peaks at $L=\pm2$~rlu correspond to H points in the next Brillouin zones. We have also analysed the triply degenerate point at the lower energy of $E=16$~meV in Supplementary Figure~5, which confirms our conclusion on the observation of two triply degenerate nodal points at H.  

Results from similar practices for the P points in Fig.~\ref{fig:dispersion}c at an energy interval of $\Delta E=15\pm0.5$~meV are presented in Fig.~\ref{fig:ecuts}b, from which we observe two Dirac nodes at the P points on top of the ring centring (-2,\,0,\,1), $i.e.$, the H$^\prime$ point. Two P and $\Gamma$ points in other directions are also present in this constant-$E$ contour. We perform a cut through the two P points along the [111] direction, and the results are shown in Fig.~\ref{fig:ecuts}d, with two peaks centring at the P points.

We have also performed a series of constant-{\bf Q} (wave-vector) cuts of the dispersions at the {\bf Q} positions indicated in Fig.~\ref{fig:qcuts}a and b through the H and P points, respectively. Results of these cuts are shown in Fig.~\ref{fig:qcuts}c-e. From the linear cuts, we identify two energies around 15 and 18~meV corresponding to these points at $q=0$ (Fig.~\ref{fig:qcuts}c) and $\mp0.5$ (Fig.~\ref{fig:qcuts}d and e), evidenced by the sharpest peaks at these Dirac and triply degenerate nodes. We also illustrate the linear dispersions near these nodes in Fig.~\ref{fig:qcuts}c-e. 

To further characterise the triply degenerate points at H, we perform simulations using the effective Hamiltonian derived from our $J_1$,$J_2$...$J_6$ model~(Eq.~\ref{H}). The results are plotted in Supplementary Figure~6, which shows that, near each of the H points, there are two linear bands and one flat band. Thus, the magnons can be regarded as three-component bosons\cite{PhysRevB.97.094412,0295-5075-120-5-57002}, similar to the new fermions\cite{Bradlynaaf5037,nature547_298,PhysRevLett.119.206402,np14_349}.
The two linear bands have a Chern number of 2 and -2, respectively, and the flat band has a Chern number of 0~(refs~\onlinecite{PhysRevB.97.094412,0295-5075-120-5-57002,Bradlynaaf5037,nature547_298,PhysRevLett.119.206402,PhysRevLett.120.016401}).

\bigskip
\noindent {\bf Discussions}\\
By now, we have unambiguously demonstrated the coexistence of Dirac and triply degenerate magnons in \cto, and so this material is the first topological system where both Dirac and triply degenerate nodal excitations are present, enabling the investigations into the possible interplay between them and other topological properties of the material. Due to the presence of the nontrivial Berry curvature in topological magnons, the anomalous thermal Hall transport resulting from the spin current is expected\cite{PhysRevLett.104.066403,Onose297,PhysRevLett.106.197202,PhysRevB.85.134411,1367-2630-18-10-103039,nc6_6805,PhysRevLett.115.106603}. 
In \cto, under an external magnetic field, a Dirac point should split into two Weyl points carrying a monopole charge of 1 and -1, respectively, which will give rise to the thermal Hall conductivity\cite{wehling2014dirac,RevModPhys.88.021004}. Furthermore, the two bands with a Chern number of $\pm2$ crossing with the flat band at the triply degenerate H point are also expected to show a thermal Hall effect under an external magnetic field that opens a gap\cite{nc6_6805}. Another important feature of the topological magnons in \cto is the topologically-protected surface arc state\cite{PhysRevLett.119.247202}, which may be detected using surface-sensitive probes, such as high-resolution electron energy loss spectroscopy\cite{doi:10.1063/1.4928215}, or helium atom energy loss spectroscopy\cite{0295-5075-4-7-013}. Recent developments in optical measurements of the spin excitations via the magneto-optical effect may also be helpful\cite{PhysRevLett.88.227201,PhysRevLett.115.197201,nc8_15859}. Furthermore, spin current flowing on the surface may be directly measured \cite{PhysRevB.97.081106}. Further explorations of these topological properties should lend support to developing spintronics with outstanding performance\cite{RevModPhys.90.015005,Onose297,PhysRevB.97.081106,chumak2015magnon}. 
Finally, since the band topology does not rely on the constituent quasiparticles, both the electron and phonon bands of \cto may exhibit topological properties\cite{Bradlynaaf5037,nature547_298}, calling for future theoretical and experimental investigations. 

\noindent {\it Note added:} After we finished this work, we became aware of a preprint reporting similar INS results\cite{arXiv:1711.00632}.

\bigskip
\noindent {\bf Methods}\\
\noindent {\bf Single-crystal growth and characterisations.} High-quality single crystals of \cto were grown using PbCl$_2$~(4N) as the flux, following the procedures in ref.~\onlinecite{HE2014146}. X-ray diffraction data were collected in an x-ray diffractometer (X$^\prime$TRA, ARL) using the Cu-$K_\alpha$ edge with a wave length of 1.54~\AA. Rietveld refinements on the data were run in the Fullprof. suite. A single crystal x-ray diffractometer was used to confirm the orientation of the single crystals. Susceptibility and heat capacity were measured in the physical property measurement system (PPMS-9T) from Quantum Design.

\medskip
\noindent{\bf Inelastic neutron scattering experiment.} Our inelastic neutron scattering experiment was performed on wide angular-range chopper spectrometer~(ARCS) at Spallation Neutron Source (SNS) of Oak Ridge National Laboratory (ORNL). For the experiment, we coaligned 40 pieces of single crystals weighed about 3~g in total using a backscattering Laue x-ray diffractometer. The single crystals glued on an aluminum plate with a sample mosaic of 1.5 degrees were loaded into a closed-cycle refrigerator with the [010] direction aligned in the vertical direction. Data were collected by rotating the sample about the [010] axis with an incident energy $E_i=35$~meV and a chopper frequency of 300~Hz resulting in an energy resolution about 1.4~meV. We collected data at various temperatures. At 5~K, the data were collected by rotating the sample by 90 degrees in a 1.25-degree step. For other high temperatures, data were collected with a 5-degree step. We used DAVE\cite{Azuah2009} to analyse the data. The wave vector {\bf Q} was expressed as {\bf Q}=$(2\pi/a,\,2\pi/b,\,2\pi/c)$ reciprocal lattice unit~(rlu) with $a=b=c=9.537(3)$~\AA. Data in Fig.~2a, b, and c were obtained by integrating the experimental data along two other orthogonal directions, with a thickness of $[H,\,0,\,0]=[-3.2,-2.8]$, $[0,\,K,\,0]=[-0.2,0.2]$; $[-L,\,0,\,L]=[0.8,1.2]$, $[0,\,K,\,0]=[-0.2,0.2]$; and $[-L,\,0,\,L]=[1.3,1.7]$, $[-K,\,2K,\,-K]=[0.1,0.2]$, respectively. Data in Fig.~3a and b were integrated over $[H,\,0,\,0]=[-3.2,-2.8]$ and $[-L,\,0,\,L]=[1.3,1.7]$, respectively.

\medskip
\noindent{\bf Linear-spin-wave theory.} In order to fit the experimental data, we used the Heisenberg model involving exchange interactions up to the sixth nearest neighbour (NN),
\begin{equation}\label{H}
  H=\sum_{n=1}^6 \sum_{\langle ij\rangle\in\{n\text{-}\rm{NN}\}} J_n {\bf S}_i\cdot{\bf S}_j,
\end{equation}
where $n$-NN indicates the $n$-th NN bond and $J_n$ is the magnitude of the $n$-th NN Heisenberg term. Since this material is a collinear antiferromagnet, we performed the calculations with the linear-spin-wave theory. After performing standard Holstein-Primakoff transformation and diagonalizing the quadric Hamiltonian, we obtained the magnon dispersions.

To compare with the experimental data, we calculated the neutron scattering cross section
\begin{equation}\label{cross}
  \frac{d^2\sigma}{d\Omega dE}\propto \sum_{\alpha\beta}(\delta_{\alpha\beta}-Q_{\alpha}Q_{\beta}/{\bf Q}^2) S^{\alpha\beta}({\bf Q},E),
\end{equation}
where $Q_{\alpha=x,y,z}$ is the $\alpha$ component of {\bf Q} and the $S^{\alpha\beta}({\bf Q},E)$ is the spin-spin correlation function defined by
\begin{equation}\label{SS}
   S^{\alpha\beta}({\bf Q},E)=\frac{1}{N}\sum_{ij}e^{\mathrm{i}{\bf Q}({\bf\rm r}_i-{\bf\rm r}_j)}\int_{-\infty}^{\infty}\left\langle S_{i}^\alpha S^\beta_{j}(t)\right\rangle e^{-\mathrm{i}Et}dt.
\end{equation}
Here, $S_i$ is the effective spin at site $i$ with the coordinate {\bf r}$_i$.

\medskip
\noindent{\bf Data Availability.} Data supporting the findings of this study are available from the corresponding author J.S.W. (Email: jwen@nju.edu.cn) upon reasonable request.



\bigskip
\noindent {\bf Acknowledgements}\\
\noindent Work at Nanjing University was supported by National Natural Science Foundation of China with Grants Nos 11674157, 11774152, 11374138, 11674158 and 11525417, National Key Projects for Research \& Development of the Ministry of Science and Technology of China with Grant No.~2016YFA0300401, and Fundamental Research Funds for the Central Universities with Grant No.~020414380105. The research at Oak Ridge National Laboratory’s Spallation Neutron Source was sponsored by the US Department of Energy, office of Basic Energy Sciences, Scientific User Facilities
Division. We thank Jian Sun, S.~A.~Owerre, Yuan Li, and Ka Shen for stimulating discussions.

\medskip
\noindent {\bf Author contributions}\\
\noindent J.S.W. and J.-X.L. conceived the project. S.B. grew the crystals. S.B. and Z.M. carried out the neutron scattering experiments with help from D.L.A.. W.W., S.-L.Y., D.W. and X. G. W. performed the theoretical calculations. S.B., J.H.W., Z.W.C. and J.S.W. analysed the data. J.S.W., J.-X.L. and S.B. wrote the paper with inputs from all authors.

\medskip
\noindent {\bf Competing Interests}\\
\noindent The authors declare no competing interests.

\end{document}